\documentclass[]{spie}  

\usepackage[]{graphicx}

\title{In-Flight Calibration of the {\it\Large Chandra} High
Energy Transmission Grating Spectrometer}

\author{Herman L. Marshall, Daniel Dewey, and Kazunori Ishibashi
\skiplinehalf
Center for Space Research, Massachusetts Institute of Technology,
Cambridge, MA 02139 \\
}

\authorinfo{Further author information: (Send correspondence to
H.L.M.)\\H.L.M.: E-mail: hermanm@space.mit.edu, Telephone: 1 617 253 8573}

\begin{document} 
  \maketitle 

\newcommand\arcdeg{\mbox{$^\circ$}}%
\newcommand\arcmin{\mbox{$^\prime$}}%
\newcommand\arcsec{\mbox{$^{\prime\prime}$}}%
\let\jnlstyle=\rmfamily
\def\refj#1{{\jnlstyle#1}}%
\newcommand\aj{\refj{AJ}}%
\newcommand\araa{\refj{ARA\&A}}%
\newcommand\apj{\refj{ApJ}}%
\newcommand\apjl{\refj{ApJ}}%
\newcommand\apjs{\refj{ApJS}}%
\newcommand\ao{\refj{Appl.~Opt.}}%
\newcommand\aap{\refj{A\&A}}%
\newcommand\mnras{\refj{MNRAS}}%
\newcommand\pasp{\refj{PASP}}%
\newcommand\pasj{\refj{PASJ}}%
\newcommand\nat{\refj{Nature}}%
\newcommand\iaucirc{\refj{IAU~Circ.}}%
\newcommand\aplett{\refj{Astrophys.~Lett.}}%

\begin{abstract}
We present results from in-flight calibration of the High Energy
Transmission Grating Spectrometer (HETGS) on the Chandra X-ray
Observatory.  Basic grating assembly parameters such as
orientation and average grating period were measured using
emission line sources.  These sources were also used to determine
the locations of individual CCDs within the flight detector. 
The line response function (LRF) was modeled in
detail using an instrument simulator based on pre-flight
measurements of the grating alignments and periods.  These LRF
predictions agree very well with in-flight observations of sources
with narrow emission lines.
Using bright continuum sources, we test the consistency of the
detector quantum efficiencies by comparing positive orders to
negative orders.

\end{abstract}



\section{Introduction}
\label{sect:intro}  

We report on several sets of observations obtained with the
{\em Chandra} High Energy Transmission Grating (HETG)
to verify and measure various properties of the system.
See Weisskopf et al. (2000,2002)\cite{weisskopf2000,weisskopf2002}
for an overview of the {\em Chandra} instrumentation and early results.
The HETG is described in detail elsewhere\cite{css86,markert94}
and in the {\em Chandra} Proposers' Observatory Guide.\footnote{Revision 5
of the Observatory Guide
was published in December 2002 and is available on-line at
{\tt http://asc.harvard.edu/proposer/POG/index.html}.}  Briefly,
the HETG Spectrometer (HETGS) operates with
the {\em Chandra} High Resolution
Mirror Assembly (HRMA), which focusses X-rays through
two sets of gratings in the HETG: high energy gratings (HEGs)
with a period of 200.081 nm and the medium energy gratings (MEGs)
with an average period of 400.141 nm.  These disperse X-rays to
the focal plane detector, which is usually the
Advanced CCD Imaging Spectrometer (ACIS)\cite{garmire2003}.

Ground-based calibration observations have been previously
reported\cite{dewey97,marshall_hetglrf,dewey98,marshall_mctest,davis98,schulz98}
and are summarized in the HETGS Ground Calibration Report\footnote{Available
at {\tt http://space.mit.edu/HETG/report.html}.}  These reports
provide detailed ground-based measurements and modeling of
various aspects of the HETGS such as
the line response function (LRF) and the grating efficiencies.
Here, we report on analysis of many observations obtained with
the HETGS during in-flight calibration and instrument checkout.

The primary intent of these observations was to verify system
performance using sources where some aspects are well understood.
Many of
the observations were obtained during the
on-orbit activation and checkout (OAC), a two-week period in August,
1999 that began just after the ACIS detector door was opened.
Others were obtained as part of a regular program to monitor the
performance of the HETGS.
Table~\ref{tb:hetg-flight-cal} summarizes calibration
data obtained and the use of
these data sets.  They are grouped according to the temperature of
the focal plane detector (ACIS) and by observing cycle

  \begin{table}
    \caption{HETGS Calibration Observations}\index{HETG!flight calibration}
    \label{tb:hetg-flight-cal}
    \centering
    \begin{tabular}{|cccc|}
\multicolumn{4}{c}{} \\  
	\hline
{\bf Obsid(s)} & {\bf Date(s)} & {\bf Target} & {\bf Comments} \\
	\hline
	\hline
\multicolumn{4}{|c|}{ACIS temperature = -100 C} \\
1098  & 8/28/99 & Capella & HETGS First-light \\
1101, 1237  & 8/29/99 & Capella & Focus set: ``+0.2 mm'' \\
1100, 1236  & 8/28/99 & Capella & Focus set: ``-0.2 mm'' \\
1099, 1235  & 8/28/99 & Capella & Focus set: ``0.0 mm'' \\
168, 169, 170  & 8/29/99 & Crab Nebula & Effective Area and Timing\\
	\hline
	\hline
\multicolumn{4}{|c|}{ACIS temperature = -110 C} \\
1103, 1318  & 9/24,25/99 & Capella & Emission Line Project \\
1102  & 9/23-24/99 & Cyg X-2 & Effective Area \\
62538, 1252 & 9/14,17/99 & HR 1099 & Emission Line Project \\
457   & 11/5/99 & Mkn 421 & Effective Area \\
459   & 1/10/00 & 3C 273 & Effective Area \\
	\hline
	\hline
\multicolumn{4}{|c|}{ACIS temperature = -120 C} \\
1705  & 5/31/00 & PKS 2155-304 & Effective Area \\
1714  & 5/29/00 & Mkn 421 & Effective Area \\
57    & 3/3/00 & Capella & Wavelength Stability \\
	\hline
1014  & 3/28,29/01 & PKS 2155-304 & Effective Area \\
1010  & 2/11/01 & Capella & Wavelength Stability \\
2463  & 6/13/01 & 3C 273 & Effective Area \\
	\hline
3167  & 11/30/01 & PKS 2155-304 & Effective Area \\
2583  & 4/29/02 & Capella & Wavelength Stability \\
3456, 3457, 3573  & 6/5/02 & 3C 273 & Effective Area and SIM offsets \\
	\hline
3706, 3708  & 11/29/02 & PKS 2155-304 & Effective Area and SIM offsets \\
4430  & 7/7/03 & 3C 273 & Effective Area \\
	\hline
    \end{tabular}
  \end{table}
%

\section{Basic Grating Parameters}
\label{sect:basics}

\subsection{Grating Focus and Dispersion Axes}

\begin{figure}
   \begin{center}
   \begin{tabular}{c}
   \includegraphics[height=12cm,angle=90]{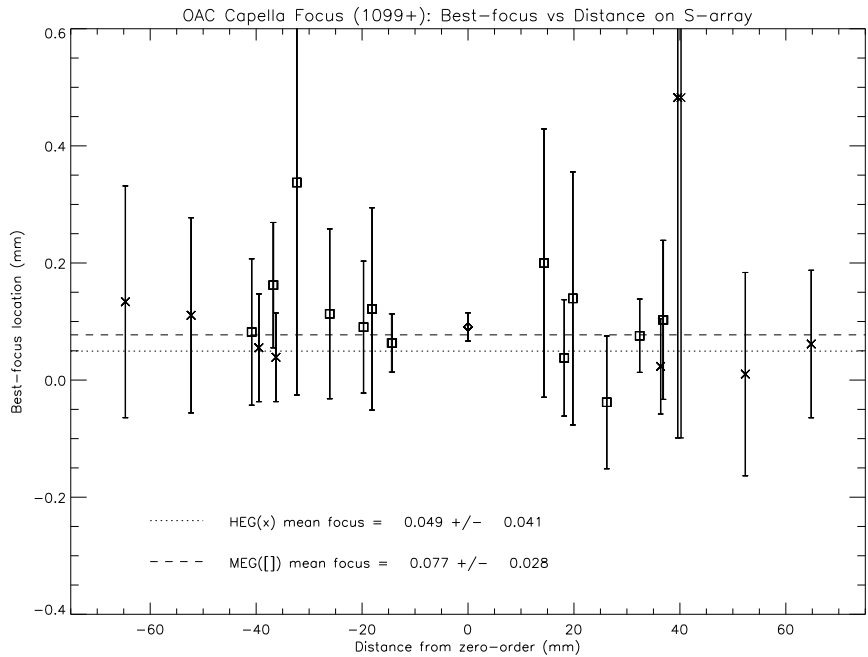}
   \end{tabular}
   \end{center}
   \caption
   { \label{fig:grating-focus}
   The best-focus location and uncertainty-range is shown for the
   zero-order (diamond) and a variety of HEG and MEG lines
   in Capella's spectra which cover a wide range along the ACIS-S detector.
   The adopted focus location of 0.054 mm is very close to the HEG
   best focus value determined here (dotted line.) }
   \end{figure}

We analyed OAC observations of Capella at three
different detector locations, $x$, along the focus direction
(see Table~\ref{tb:hetg-flight-cal}) in order
to locate the best focus, $\hat{x}$.
The full width at half-maximum (FWHM) of an emission line
along the dispersion direction varies quadratically with $x$:

\begin{equation}
\label{eq:wvsdx}
{\rm FWHM}^2(x)  = {\rm FWHM}_{\rm min}^2 + C (x - \hat{x})^2
\end{equation}

\noindent
where $C$ is a constant related to the spectrometer geometry
and FWHM$_{\rm min}$ depends on the quality of the optics.
Using eq.~\ref{eq:wvsdx}, we obtained
$\hat{x}$ for each line and the zeroth order
(Fig.~\ref{fig:grating-focus}).
The zeroth order images in the data are heavily affected by
pileup and not useful.  Instead, the ``trailed'' (readout
streak) events of the zeroth order were used.
The best focus
was +0.077 $\pm$ 0.028 mm (+0.049 $\pm$ 0.041 mm)
for the dispersed MEG (HEG) emission lines.
Thus, the overall HETG best-focus is in range of 0.050 to 0.100 mm
with respect to the initial reference point (at step -505 of
the Science Instrument Module [SIM] focus motor).
Separate analyses by the imaging team yielded a best ACIS-S focus
at 0.054 mm, so it was convenient and
reasonable to set the HETGS focus at this same value (motor step -468).

The angles of the HEG and MEG spectra on the ACIS detector were
measured using the Fe {\sc XVII} 15 \AA\ line, which is
very bright in the observations of Capella,
(observation IDs 1099 and 1235).
The angles were measured as the slope of the line
connecting the centers of the plus and minus first-order images of the
line, giving $4.725 \pm 0.01$\arcdeg\ for the MEG and
$-5.235 \pm 0.01$\arcdeg\ for the HEG.  The angle between the dispersion
lines is $9.96 \pm 0.01$\arcdeg, which is consistent with the ground-based
measurement of $9.93 \pm 0.01$\arcdeg.

\begin{figure}
   \begin{center}
   \begin{tabular}{c}
   \includegraphics[height=12cm,angle=90]{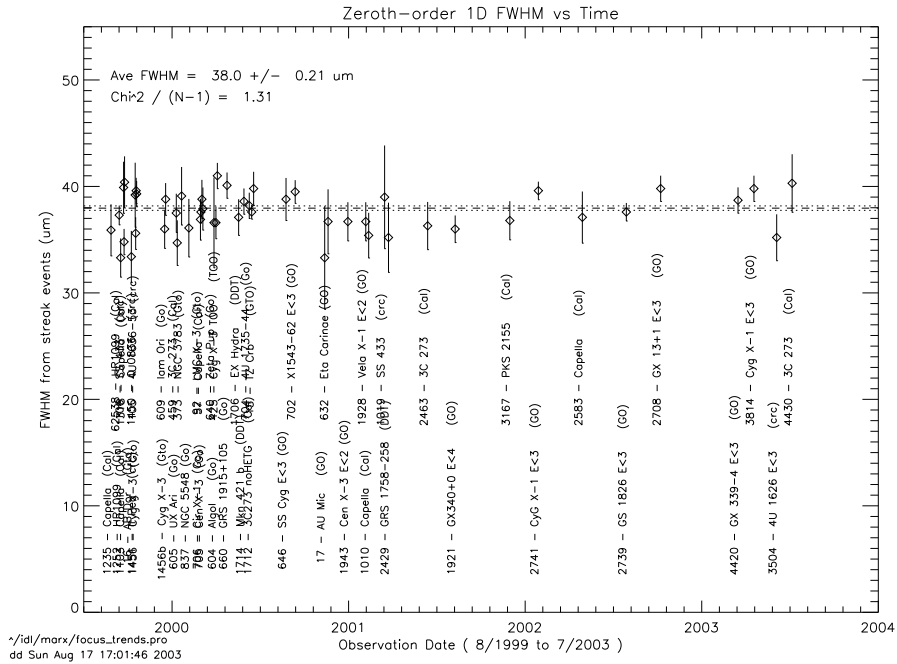}
   \end{tabular}
   \end{center}
   \caption
   { \label{fig:focus-time}
The full width at half-maximum (FWHM) of the readout streak for many
bright sources as a function of observation date.  For bright sources,
the image is often heavily affected by pileup, so the readout streak
is used, providing a 1D profile of the point spread function.  The
data are tightly distributed about an average of 38 $\mu$
($\approx$0.8\arcsec), showing that the focus has been
stable for four years.
}
   \end{figure}

The FWHM of the 1D projection of the zeroth order readout
streak for HETGS observations has been monitored for about four
years.  The width of the PSF can be broader for sources with
harder spectra, so filtering on $E<$2 keV or $E<$3 keV has been applied
in these cases, in order to provide a better comparison to other,
softer sources.  From Fig.~\ref{fig:focus-time}, it
is clear that the FWHM has been constant
over this long baseline (with SIM focus motor still set at step -468).
Note that these analyses included pixel randomization in the ACIS
processing; the average FWHM value of 38.0 $\mu$m determined here is
expected to be $\approx$34 $\mu$m when pixel randomization is eliminated.

\subsection{Wavelength Accuracy}

Relative wavelength accuracy is affected by the knowledge of the
positions of the chips in the ACIS detector and accuracy of
the dispersion scale, which is set by the HEG and MEG grating periods
and the distance from the grating assembly to the focal plane.
Lines are measured relative to zeroth order,
taking out the uncertainties in the spacecraft attitude and
telescope boresight.

\subsubsection{ACIS Chip gaps}

The ACIS-S detector is made up of six independant CCD detectors
mounted in a line with approximately equal spacing.
The geometry of each CCD is 
very well defined and stable but the five chip-to-chip offsets
or gaps required calibration.
To measure the gaps, we assume that an
emission line observed on one side of the ACIS-S array
should be measured to have the same wavelength when detected on the other
side, that is $\lambda^+ = \lambda^-$.  There is no assumption
about the absolute line wavelength for this analysis, so even
blended lines may be used.  Any difference between
these wavelengths is assumed to be due to a calibration error in the
locations of the ACIS chips.
Another term that is also included allows that
the zeroth order for the HEG and MEG gratings (that is the corresponding
HRMA shell pairs) may be offset from each other.

These offsets were measured in flight using emission lines
from Capella, which is
frequently observed for in-flight calibration.
For each given line, some linear combination of the six parameters (5
chip gap values and one HEG-MEG zeroth order offset) contributes to the
difference between the measured wavelengths of the
$+1$ and $-1$ orders.  Lines are observed on many different chip combinations
so all chip offsets can be determined.  Six lines from the MEG spectra and
five HEG lines were used: Si {\sc xiii} at 6.6477\AA, Mg {\sc xii} (8.419 \AA),
Mg {\sc xi} (9.1685 \AA), Ne {\sc x} (12.132 \AA), Fe {\sc xvii} (15.013 \AA), and
O {\sc viii} (18.967 \AA).  Singular value decomposition was used to solve
this over-constrained set of linear equations (11 measurements for
6 parameters.)

Relative to the nominal design location of the CCDs, the in-flight
determination of the chip gaps in September
1999 produced offsets to the chip locations
in the ``tiled detector'' $x$ (TDETX) coordinate
of -4.97, -3.58, -2.42, 0, -0.08, and +0.33
pixels, for chips S0 to S5 respectively, where a pixel is about 0.024 mm.
(Note that the offset of chip S3 is fixed at 0 by definition.)
Capella was observed most recently in April 2002 
and analyzed in a similar fashion
to test for chip location changes.  The analysis
gave residual offsets from the above values,
0.53, 0.31, 0.20, 0, -0.03, and -0.23 pixels and an HEG-MEG
zeroth-order offset value of 0.15 pixels.
Using these changes to the offsets
the overall wavelength residuals were of order 0.05 pixel (1$\sigma$).
This shows that the chip offsets are stable and that
re-analysis of the set of Capella observations could
yield an offset calibration of order 0.1 pixel.

For reference, we note that the chip positions
derived from pre-flight metrology indicated
a set of offsets of -2.4, -2.1, -1.8, 0, -0.75, and -1.2 pixels.
Later the locations were measured
during the so-called molecular contamination
tests taken when the
telescope and flight ACIS detector were calibrated at the
Marshall Space Flight Center's X-ray Calibration Facility;
numerous emission lines due to contaminants on the C anode were
measured.
(Other results from these tests were reported by Marshall et
al. 1998\cite{marshall_mctest}.)
Relative to ACIS chip S3,
these tests yielded offsets of -3.4, -2.5, -1.9, 0, -0.3, and -0.6 pixels.
We assign an uncertainty of 0.5 pixels due to chip location uncertainties.

\subsubsection{Dispersion Scale and Overall Accuracy}

\begin{figure}
   \begin{center}
   \begin{tabular}{cc}
   \includegraphics[height=8cm]{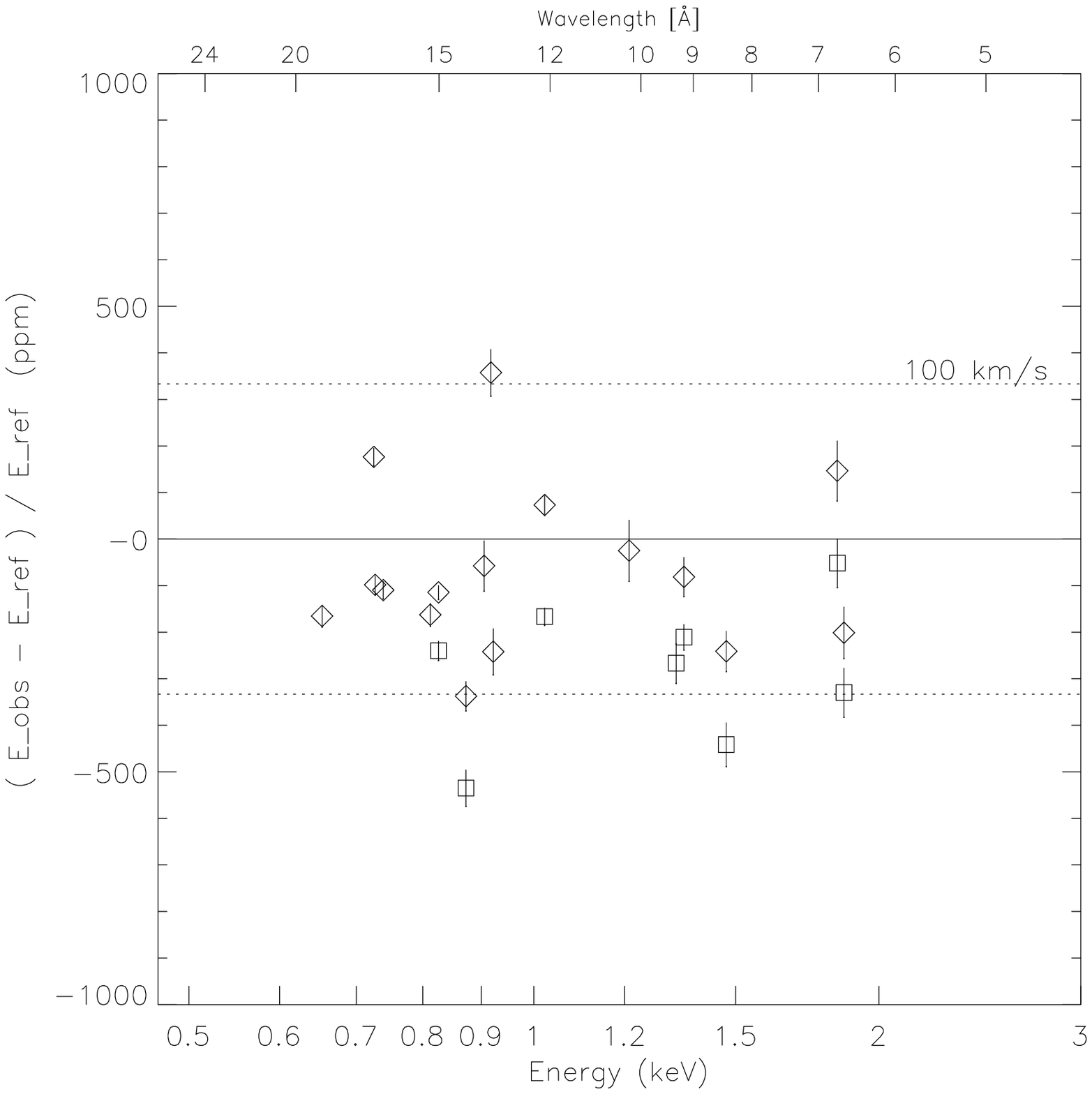}
   \includegraphics[height=8cm]{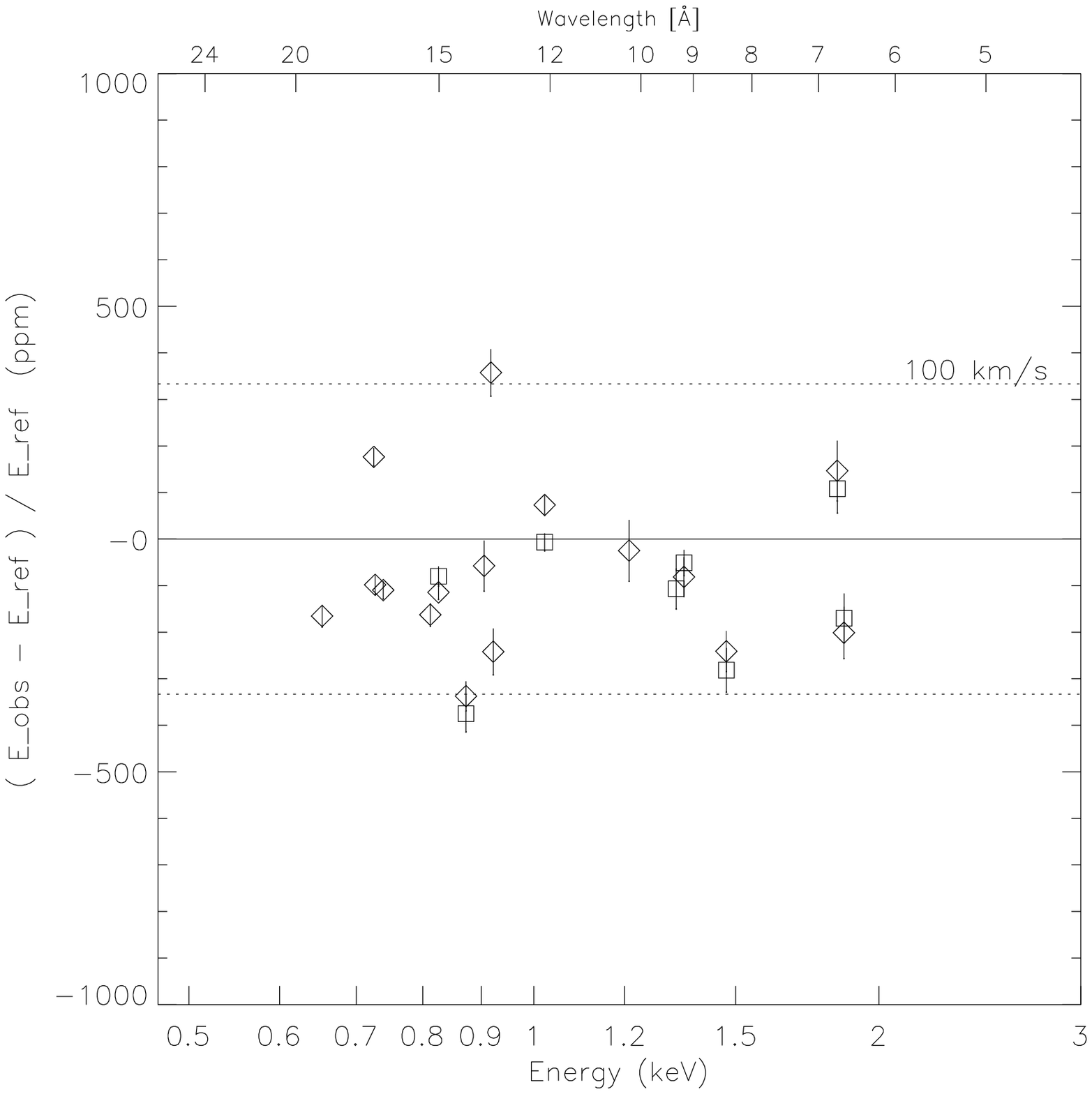}
   \end{tabular}
   \end{center}
   \caption
   { \label{fig:dispersion}
{\bf a)} (Left) Measurements of lines in one observation of Capella
(observation ID 2583).  {\em Open squares:} Lines from the HEG portion
of the spectrum.  {\em Open diamonds:} Lines from the MEG portion of
the spectrum.
{\bf b)} (Right) Same as fig.~\ref{fig:dispersion}a) except that a
correction of -0.016\% has been applied to the HEG
period, bringing the HEG wavelengths into better agreement
with the MEG wavelengths.  After correction, the average wavelength
error would result in velocity errors of less than 100 km/s.
}
   \end{figure}

After correcting for chip location errors,
the absolute accuracy of wavelengths based on
preflight instrument parameters was found to have a systematic
fractional error of order 0.05\% in the sense that measured
wavelengths were too large. This difference was traced
to a reduction in the ACIS-S pixel size from 24.000 microns to
23.987 microns as the detector
substrate was cooled to the in-flight operating
temperature of -120 C. This reduction was verified
in preliminary analysis of calibration observations of the star
cluster NGC 2516 by Maxim Markevitch. (The chip gap offsets, above,
are unaffected by this scale change
because they were measured using in-flight data.)
After correcting for this scale error, MEG wavelengths show
no systematic deviations from their
expected values (Fig.~\ref{fig:dispersion})
but the HEG wavelengths still show a systematic shift.  This
shift can be corrected by rescaling HEG wavelengths by
-0.016\%, corresponding to an HEG period of 200.049 nm.
Thus, after applying these corrections, we estimate that the
relative accuracy of wavelengths assigned to events on the
basis of the distance from zeroth order appear to be good to better
than 0.0055 \AA\ (0.0028 \AA) for MEG (HEG) spectra, limited
primarily by the uncertainties in the ACIS chip locations.

\section{HETGS Line Response Function}
\label{sect:lrf}

The HETGS Line Response Function, LRF, consists of a Gaussian-like core
and extended wings.  The resolution of the spectrometer is specified by
the FWHM of the core; the core plus wings are relevant to accurate
line flux measurements.  These properties are captured as resolution (or
resolving power, $E/dE$) values and RMF files.

\begin{figure}
   \begin{center}
   \begin{tabular}{c}
   \includegraphics[height=12cm,angle=90]{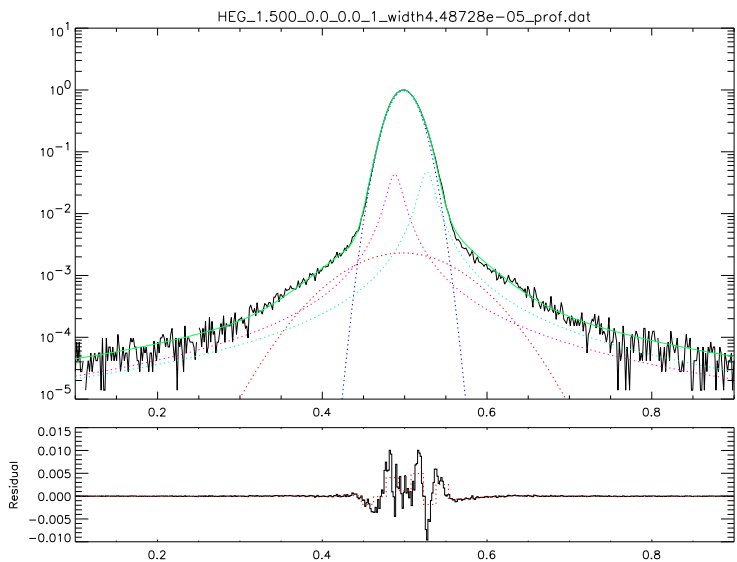}
   \end{tabular}
   \end{center}
   \caption
   {
Example of a fit to an emission line generated using
the {\tt MARX} simulation software.  The input line energy was
1.500 keV and only the HEG $+1$ order is shown.  The line response function 
is modeled by two Gaussian components and two Lorenztian components.
The narrow Gaussian dominates the total power.
The Lorentzian wings are rarely detectable in flight data.
The residuals (bottom panel) are largest in the core but are
not statistically significant.\label{fig:lrf-example} }
   \end{figure}

\begin{figure}
   \begin{center}
   \begin{tabular}{cc}
   \includegraphics[height=8cm,angle=270]{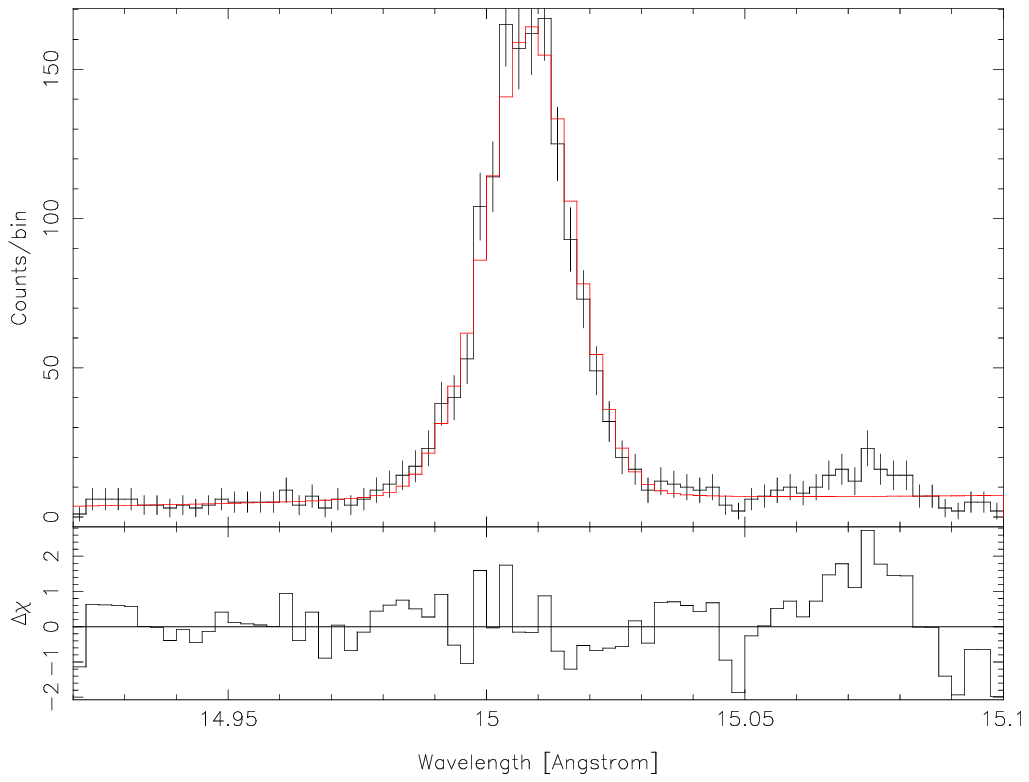}
   \includegraphics[height=8cm,angle=270]{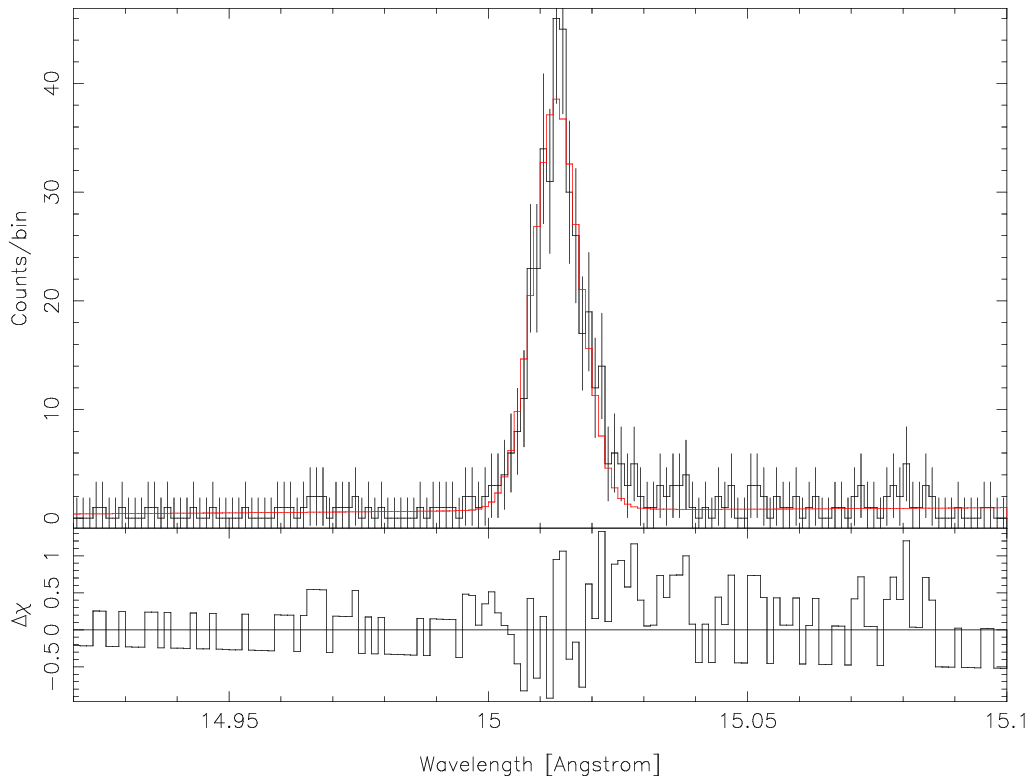}
   \end{tabular}
   \end{center}
   \caption
   { \label{fig:fe17}
Comparison of data from Capella for the Fe {\sc XVII} line at
15.013 \AA\ to the LRF model fitted by the 4 component model
shown schematically in Fig.~\ref{fig:lrf-example}.  The left
side compares the MEG data to the LRF model while the right
side shows the same thing for the HEG data.
}
   \end{figure}

\begin{figure}
   \begin{center}
   \begin{tabular}{cc}
   \includegraphics[height=8cm,angle=270]{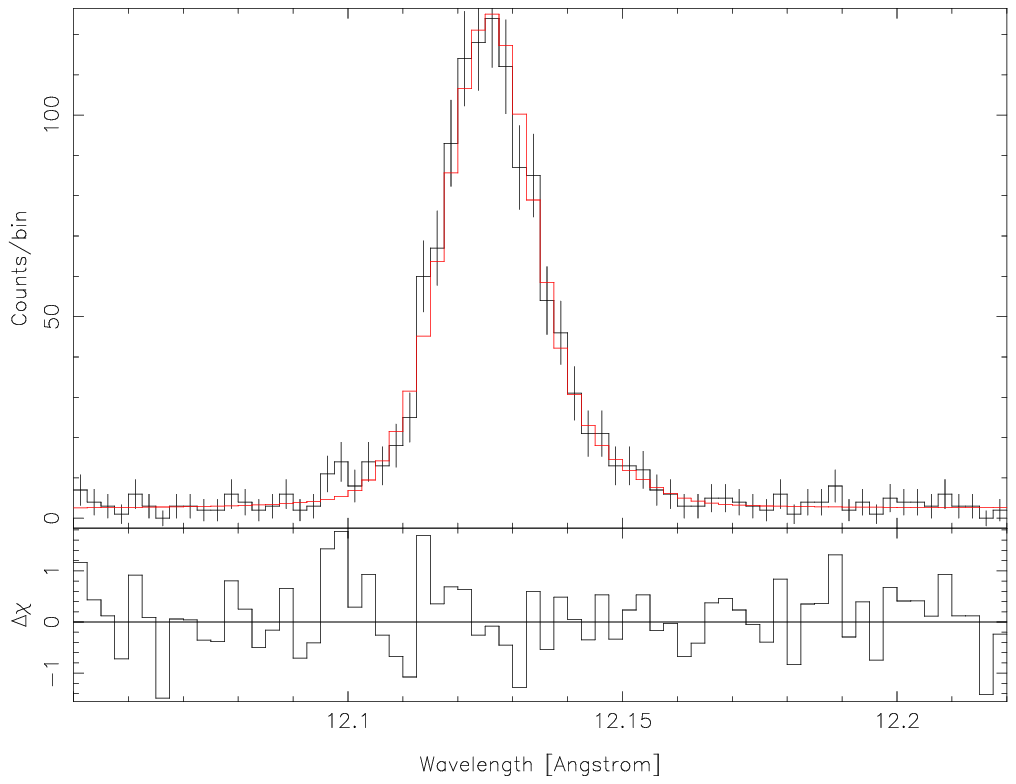}
   \includegraphics[height=8cm,angle=270]{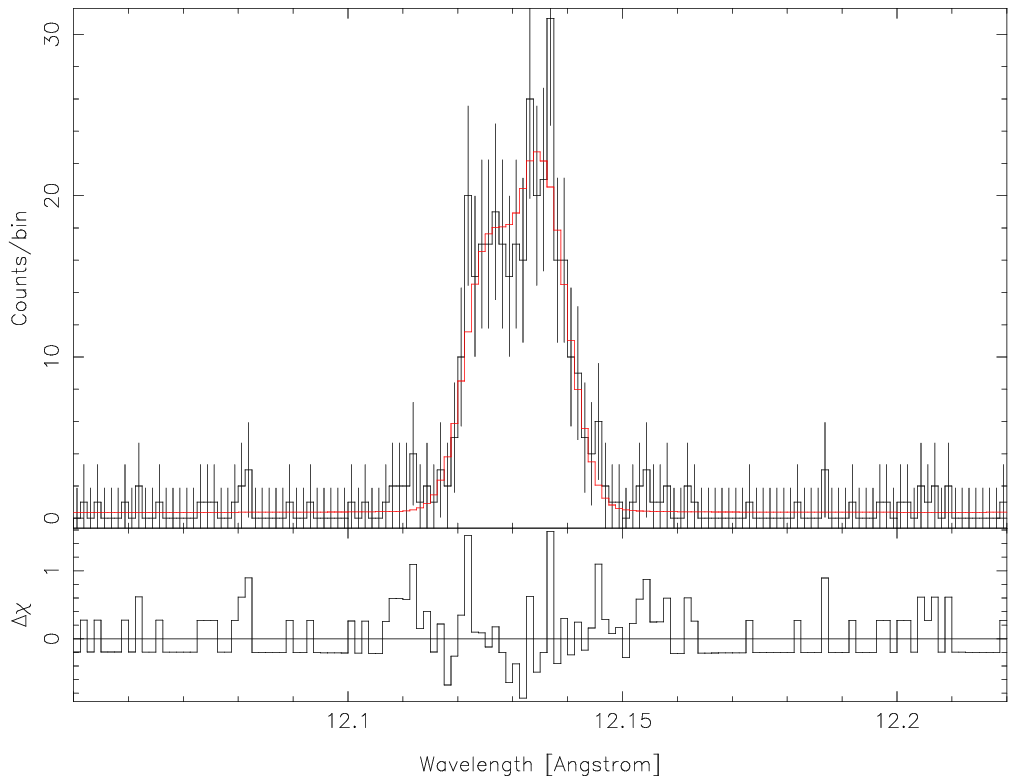}
   \end{tabular}
   \end{center}
   \caption
   { \label{fig:blend}
Same as in fig.~\ref{fig:fe17} but for a blend of two
lines.  One is Fe {\sc XVII} at 12.125 \AA\ and the
other is Ne {\sc X} at 12.136 \AA.  The two lines are
separated by 0.009 \AA; they are
not resolved by the MEG with an expected resolution
of 0.018 \AA\ at 12.13 \AA\ but are marginally resolved by
the HEG, for which the resolution is 0.0101 \AA\ (using
eq.~\ref{eq:fwhm}) demonstrating that {\tt MARX} provides
very good simulations of the HETGS LRF.
}
   \end{figure}

The results of ground testing of the HETG are incorporated into the
ray-trace simulator, {\tt MARX}\cite{marx}.
The simulation software embodies much of the ground-calibration data, so
line response functions are generated with millions of events and fitted
with a model consisting of two Gaussian components and two Lorenztian components.
The narrow Gaussian dominates the total power.
The Lorentzian wings are rarely detectable in flight data, so they
are fit to simulated data.
The simulated event list was processed
and fit using {\tt IDL}$^{\rm TM}$.
The result is the expected HETGS LRF for a point source.
An example of a very good fit is shown in fig.~\ref{fig:lrf-example}.
The FWHM is well approximated by a cubic function of the wavelength:

\begin{equation}
\label{eq:fwhm}
{\rm FWHM}  = c_0 + c_1 \lambda + c_2 \lambda^2 + c_3 \lambda^3
\end{equation}

\noindent
where $c_i$ are given in table~\ref{tab:fwhm} for the MEG and HEG
separately.  The LRFs derived from the fits match the observed data
from Capella (observation ID 1103) very
well, as shown in figs.~\ref{fig:fe17} and \ref{fig:blend}.  More details are
available on-line at {\tt http://space.mit.edu/CXC/LSF/LSF\_0002/}.

  \begin{table}
    \caption{Coefficients for FWHM as a
    	Function of $\lambda$ (Eq.~\ref{eq:fwhm})}
    \label{tab:fwhm}
    \centering
    \begin{tabular}{|lcccc|}
	\hline
{\bf Grating} & {\bf $c_0$} & {\bf $c_1$} 	& {\bf $c_2$} 	& {\bf $c_3$} \\
	\hline
	\hline
MEG & 0.018851868 & -0.00033946575  & 2.9781558e-05 & -4.1097023e-07 \\
HEG & 0.010019662  & -0.00014101982  & 1.4591413e-05  & -1.9632994e-07 \\
	\hline
    \end{tabular}
  \end{table}

\section{HETGS Effective Area}
\label{sect:area}

There are many components to the HETGS effective area model,
specifically the HRMA, HETG, and ACIS.  The focus here is on
to ACIS-related effects that contribute important residuals to
HETGS effective area calibration at this time.
Using grating observations, we may test the validity of several
components of the system without regard to the nature of the
sources.

\subsection{Comparing the BI and FI Quantum Efficiencies}

Eighteen LETGS or HETGS observations were used to compare the
quantum efficiencies (QEs) of backside-illuminated (BI) chips
relative to frontside-illuminated (FI) chips.
The observations were grouped approximately by
date centered about January 2000, June 2000, January 2001, June 2001,
and November, 2001.  The data were extracted from L1 event lists
processed using custom IDL scripts that have been used for other
observations \cite{marshall01,marshall02,jem01}.
Briefly, the extraction region is 0.001\arcdeg\ wide and the
ACIS pulse height data were used to select events in wide regions that
accept 95-99\% of the events.  All observations were obtained using ACIS
in timed exposure (TE) mode except for one LETGS observation in
continuous clocking (CC) mode.  For the TE mode observations background
was selected from cross dispersion regions 0.003-0.01\arcdeg\ from the
dispersion line.  For CC mode, background was taken from a pulse height
region with equivalent range of $E_{ACIS} / E_{OTG}$.

The events identified with $+1$ and $-1$ orders were binned to about half of
a spectral resolution element: 0.005 \AA\ (HEG), 0.010 \AA\ (MEG), and 0.025
\AA\ (LETG).  Background was binned similarly and subtracted after
weighting by the ratio of the cross dispersion extraction widths.
Uncertainties were computed from Poisson statistics for each bin.  For
each group of observations with net counts $C_i^+$ in the $+1$ order and $C_i^-$
net counts in the $-1$ order and each grating, the ratio

\begin{equation}
R = \frac{Q^-}{Q^+} \frac{\sum_i C_i^+}{\sum_i C_i^-}
\end{equation}

\noindent
and its statistical uncertainty $\sigma_R$ were computed over adaptively
sized wavelength bins.
Note that here we are assuming equality of plus and minus order efficiency
for the gratings which is reasonable from ground tests and the random
$\pm$180\arcdeg\ installation of grating facets in the HETG.
The detector QEs on the $+1$ and $-1$ sides
$Q^+$ and $Q^-$, are
derived from the models of the CCDs in the detector and depend on
wavelength. The index $i$ indicates an individual observation in the
group.  Assuming that the QEs are not perfectly known, we assign $\hat{Q}^+$ and
$\hat{Q}^-$ to represent the {\em true} QEs on these sides, so that the expected
counts in each wavelength bin on the $+1$ and $-1$ sides are given by

\begin{equation}
C_i^+ = n_i A t \epsilon T \hat{Q}^+ d\lambda
\end{equation}

\begin{equation}
C_i^- = n_i A t \epsilon T \hat{Q}^- d\lambda
\end{equation}

\noindent
where $n_i$ is the source flux in
ph cm$^{-2}$ s$^{-1}$ \AA$^{-1}$, $A$ is the effective area of
the HRMA, $t$ is the observation exposure, $\epsilon$ is the grating efficiency
into first order, $T$ is the transmission of the detector filter, $Q^+$ ($Q^-$) is
the CCD efficiency on the $+1$ ($-1$) order side, and $d\lambda$ is the
wavelength interval corresponding to one bin.  All quantities are
functions of wavelength except $t$.  As defined, the quantity $R$ is

\begin{equation}
\label{eq:pmratio}
R = \frac{Q^-}{Q^+} \frac{\hat{Q}^+}{\hat{Q}^-}
\end{equation}

\noindent
independent of the true and perhaps unknown source model, grating
efficiency, or filter transmission.  The bin wavelength limits, $\lambda_1$
to $\lambda_2$, were limited by $R/\sigma_R < 40$ and
$\lambda_2 < 0.07 \lambda_1$.
In practice, there are very few observations for which the signal/noise
is more than 40 over a small wavelength range.  The bins have different
sizes for each grating type.

\begin{figure}
   \begin{center}
   \begin{tabular}{c}
   \includegraphics[height=17cm]{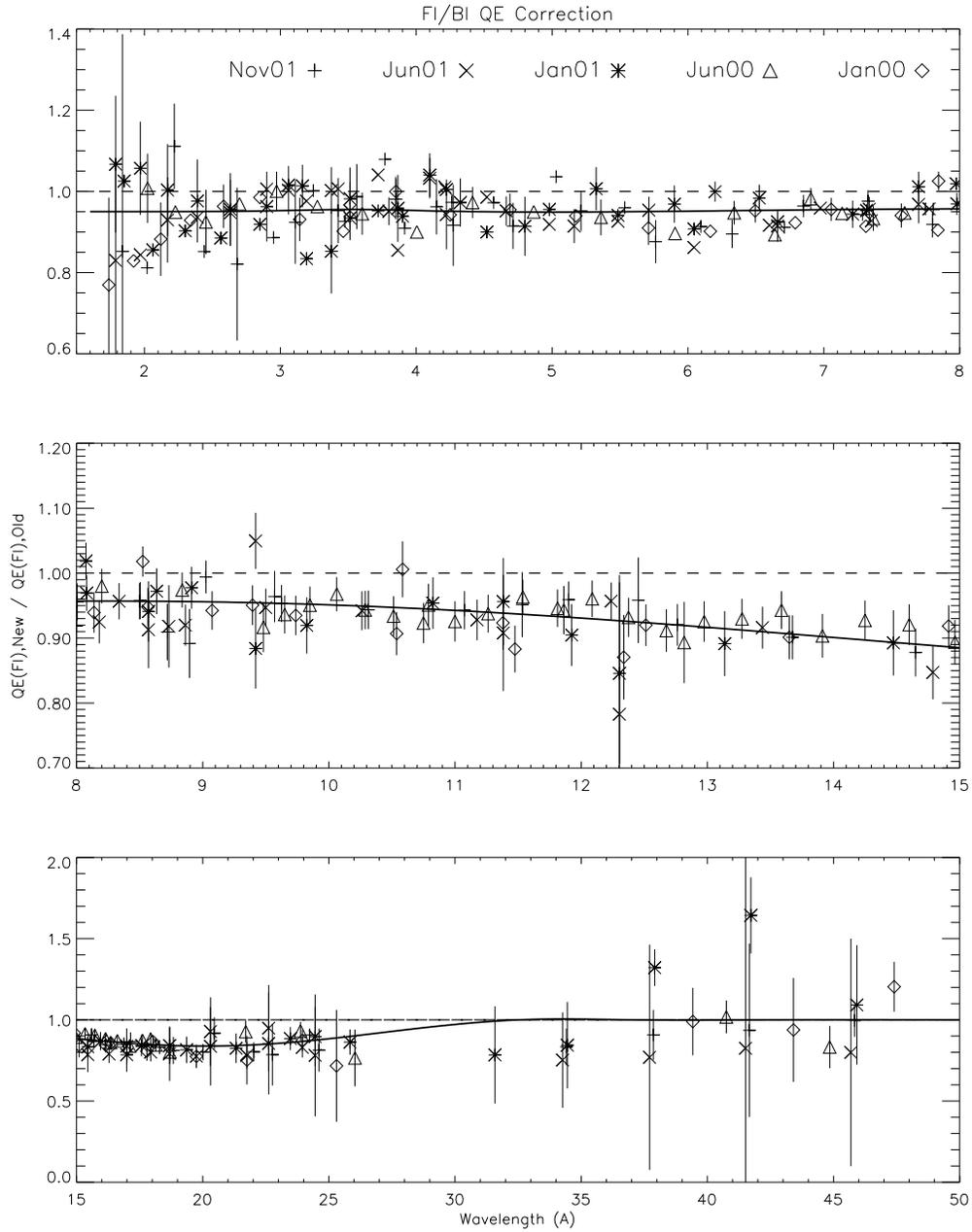}
   \end{tabular}
   \end{center}
   \caption
   { \label{fig:bifi-ratio}
The ratio $r$ as a function of wavelength
for each observation set derived using eq.~\ref{eq:bifiratio}.
The data are not segregated by grating type
because the results are independent of the grating where comparisons are
possible.  All comparisons involving two FI chips gave $R$
(eq.~\ref{eq:pmratio}) consistent with
1 and are not shown.  The heavy line is a polynomial fit to the data
with Gaussian roll-off to the ends to avoid poor behavior of the
polynomial extrapolation where the data are not as good. }
   \end{figure}

The measurements of $R$ are separated according to the type of CCD on the
$+1$ and $-1$ sides to form a physically useful ratio

\begin{equation}
\label{eq:bifiratio}
r = \frac{Q_{BI}/Q_{FI}}{\hat{Q}_{BI}/\hat{Q}_{FI}}
\end{equation}

\noindent
which gives the ratio of true QEs relative to the modeled QEs.  If we
have reason to believe that the the BI QEs are correct, for instance,
then $r$ gives the correction factor by which we multiply the modeled FI
QEs to obtain the {\em true} QEs.  The results are shown in
fig.~\ref{fig:bifi-ratio}.

\begin{figure}
   \begin{center}
   \begin{tabular}{c}
   \includegraphics[height=10cm]{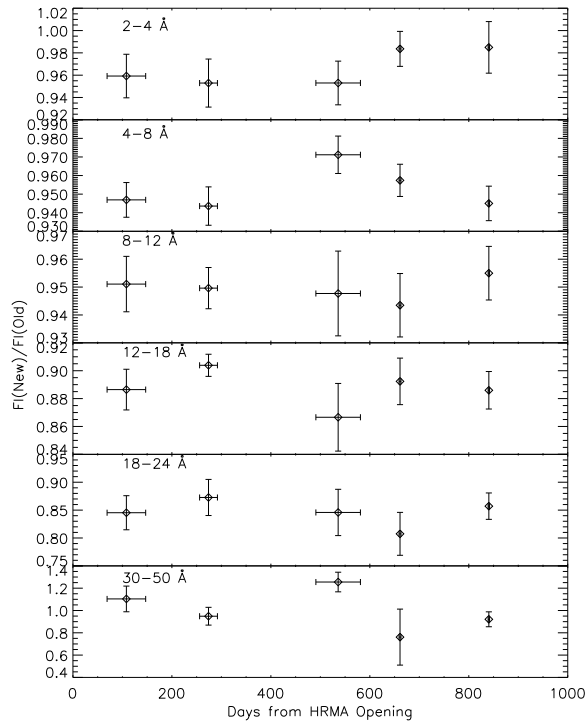}
   \end{tabular}
   \end{center}
   \caption
   { \label{fig:bifi-time}
The data in fig.~\ref{fig:bifi-ratio} were averaged over
large wavelength regions due to the weak wavelength dependence and shown
here as a function of observation time.  The horizontal bars represent
the time range of the observations within each group.  These data are
consistent with no change of the ratio of the QEs as a function of time.
Limits to the secular trends have not been estimated but are rather
small. }
   \end{figure}

To check for time dependence, several wide wavelength intervals
were averaged where $r$ changes slowly.  The ratios from different
observations in similar time intervals were combined, where the
time interval did not exceed 6 months.  The results are shown
in fig.~\ref{fig:bifi-time}.
Limits to secular drifts are relatively
stringent near 1 keV, $<$ 1\% change per year, and are somewhat larger at
longer wavelengths: $<$ 10\% per year in the 20-25 \AA\ range
and $<$ 30\% per year in the 30-50 \AA\ range.

\subsection{Accounting for Modest Pileup}

Pileup is a highly non-linear effect that reduces the total counts
selected for any specific grating order.  Comparing the MEG and
HEG spectra of the Crab pulsar shows what might happen.
See the Proposers'
Observatory Guide section 8.2.1.4 (particularly Fig. 8.14) for a
detailed description of the effects of pileup in HETGS spectra.
An empirical approach was used to make a first-order correction for
pileup that reduces the effect of the Ir-M edge to a tolerable
level.

A simple correction factor is applied to the effective areas based
on the observed count spectrum.  The event list is binned in the
spectral dimension at a wavelength interval corresponding to one
ACIS pixel (which depends on the grating), giving $C_i$, the counts
in wavelength bin $i$.  The total number of frames, $N_f$, is
determined from the observation length and the frametime (less the
framestore transfer time), giving the rate, $\rho_i = C_i/N_f$, in
counts/frame in any given wavelength bin.  The effective area
correction is then

\begin{equation}
A^{\prime} = A e^{-a \rho_i}
\end{equation}

\noindent
for an FI chip or

\begin{equation}
A^{\prime} = A e^{-b \rho_i - c \rho_i^2}
\end{equation}

\noindent
for a BI chip.
The coefficients $a$, $b$, and $c$ were determined using an observation
of XTE J1118-480 to be 7.5, 6.38, and 22.92.  Note that the
effective area is always reduced and that the correction factor
drops rapidly with high count rates.

On a practical note, whenever there are few counts, the correction
factor is discrete and the resultant correction factor will add
noise.  One way to avoid this problem is to use an averaging
filter to determine the count rate for any given wavelength bin.

This approach has only been tested and seems to be effective for
cases of mild to moderate pileup: rates less than about 0.1-0.2
count/frame/bin.  For more extreme cases, a more detailed model is
necessary.  One method is described by Davis
(2001,2003)\cite{davis01,davis03}.\footnote{See also the web page
{\tt http://space.mit.edu/CXC/analysis/PILECOMP/index.html}.}

\subsection{Overall Effective Area Reliability}

Based on the results from the HETGS calibration observations of the Crab
pulsar and Mk 421 we place the following estimates on the
systematic uncertainties of HETGS spectral fluxes, after correcting for
the system effective area using the currently released calibration data
products: 10\% for $1.5 < E < 6$ keV (both MEG and HEG), 20\% for $6 < E < 8$
keV (HEG only), 20\% in the Si-K edge region (1.83-1.84 keV) (both MEG
and HEG), 20\% for $0.8 < E < 1.5$ keV (both MEG and HEG), 30\% for $0.5 < E
< 0.8 keV$ (MEG only), 50\% in the O-K edge region (0.525-0.57 keV) (MEG
only).
Except for the last two regions, one may conservatively use 10\%
for $E > 2$ keV and 20\% for $E < 2$ keV except in the edge regions.
The MEG
is probably better than 20\% in the 0.5-0.8 keV region but the data are not
conclusive.
The quoted uncertainties are meant to give the uncertainty
on fluxes determined in one band relative to those in another band.
Absolute flux uncertainties have not yet been verified but are expected
to be good to a better than 5-10\%.
The uncertainties include systematic
errors in many instrument components, including the HETGS efficiencies
and ACIS filter and QE models, and are meant to reflect the system
performance.

\section{Summary and Future Work}
\label{sect:summary}

On the whole, the HETGS performs as expected in flight and
in-flight calibrations are merely fine adjustments to ground-based
initial values.
In this paper we've presented some of the more important and
unique HETGS calibration activities carried out using flight data.
In many cases, these calibrations have been captured in the 
calibration data products used by processing software.  In several
cases these calibrations are in process and will show up
in the calibration products in the near future.

\acknowledgments     
 
We acknowledge and thank our many Chandra collegues.
This work was supported in part by the Smithsonian Astrophysical
Observatory (SAO) contract SVI-61010 for the Chandra
X-Ray Center (CXC) and by NASA through contract NAS8-01129.


\bibliography{hetgcal}   
\bibliographystyle{spiebib}   

\end{document}